\documentclass[a4paper]{iopart}

\usepackage{helvet}
\usepackage{mathptmx}
\usepackage[utf8]{inputenc} 
\usepackage{graphicx}
\usepackage{epstopdf}
\usepackage[
,textwidth=15cm
,textheight=20cm
,verbose
,dvips
]{geometry}
\usepackage{cite}

\begin{document}

\title{Current path in light emitting diodes based on nanowire ensembles}
\author{F Limbach, C Hauswald, J Lähnemann, M Wölz, O Brandt, A Trampert, M~Hanke, U Jahn, R Calarco, L Geelhaar, and H Riechert}
\address{Paul-Drude-Institut für Festkörperelektronik, Hausvogteiplatz 5--7, 10117 Berlin, Germany}
\ead{geelhaar@pdi-berlin.de}

\date{\today}

\begin{abstract}
Light emitting diodes (LEDs) were fabricated using ensembles of free-standing (In,Ga)N/GaN nanowires (NWs) grown on Si substrates in the self-induced growth mode by molecular beam epitaxy. Electron beam induced current analysis, cathodoluminescence as well as biased $\mu$-photoluminescence spectroscopy, transmission electron microscopy, and electrical measurements indicate that the electroluminescence of such LEDs is governed by the differences in the individual current densities of the single-NW LEDs operated in parallel, i.\,e. by the inhomogeneity of the current path in the ensemble LED. In addition, the optoelectronic characterization leads to the conclusion that these NWs exhibit N-polarity and that the (In,Ga)N quantum well states in the NWs are subject to a non-vanishing quantum confined Stark effect.

\end{abstract}

\submitto{\NT}


\newpage
\section{Introduction} 
Solid state lighting offers the promise of significant savings in electricity consumption for general lighting purposes \cite{Humphreys2008}. For this vision to come true, further improvements in the efficiency of light emitting diodes (LEDs) and, more importantly, drastic reductions in fabrication costs are required. In this context, LEDs based on group-III-N nanowires (NWs) have recently attracted more and more interest because such devices exhibit several conceptual advantages over conventional planar structures \cite{Kikuchi2004, Kim2004, Hersee2009, Bavencove2010, Guo2010, Armitage2010, Lin2010, Nguyen2011, Bavencove2011, Hahn2011, Kunert2011, Riechert2011, Waag2011, Kishino2012, Li2012}. First, in the NW geometry, strain induced by lattice mismatch can elastically relax at the free sidewalls \cite{Bjork2002}, and thus GaN NWs can be grown in excellent crystal quality on cost-effective Si substrates \cite{Geelhaar2011}. Second, for the same reason the strain in axial (In,Ga)N/GaN quantum wells (QWs) is reduced, which in turn decreases the quantum confined Stark effect (QCSE) and enhances the internal quantum efficiency \cite{Renard2009, Lahnemann2011}. Third, by growing radial core-shell QWs on the NW sidewalls, the area of the active region is increased \cite{Waag2011, Li2012}. Fourth, light extraction is expected to be higher from a NW ensemble than from a planar layer \cite{Henneghien2011, Kim2004}.

For general lighting purposes, NW-LEDs are only relevant  if they are based on ensembles. Such LEDs have been grown by metal-organic hydride vapor phase epitaxy \cite{Kim2004}, by hybrid chemical vapor deposition \cite{Hahn2011}, by metal-organic vapor phase epitaxy \cite{Hersee2009}, and by molecular beam epitaxy (MBE) \cite{Kikuchi2004, Bavencove2010, Guo2010, Armitage2010, Nguyen2011, Bavencove2011, Lin2010, Kunert2011}. Naturally, for devices based on ensembles of nanostructures the question of homogeneity arises. Both variations in electroluminescence (EL) wavelength and intensity have been reported \cite{Bavencove2011, Lin2010}. The wavelength variation has been associated with fluctuations in In content and is also observed in photoluminescence (PL) wavelength \cite{Lin2010}. With respect to EL intensity, Bavencove \textit{et al.} \cite{Bavencove2011} observed a very dramatic effect in that only about 1\% of all NWs exhibited EL at all. Obviously, the number of electroluminescent NWs has to be drastically increased for applications. However, the analysis of this observation is not straightforward because the EL intensity is not only influenced by the active region and its radiative recombination efficiency but also by the current that actually flows through the active region. Thus, NWs exhibiting very bright PL may not emit EL at all if current is not injected into them, be it due to failed contacting or significant NW-to-NW variations in series resistance. To disentangle both of these aspects is a challenging task.

Here, we analyze the current path in such LEDs by combining several characterization techniques. We show that even if all the NWs of the ensemble are capable of emitting light and are contacted, fluctuations in the series resistances of the individual NWs prevent many NWs from emitting EL. Therefore, the homogeneity of the current path is even more crucial than the homogeneity of the individual active regions.

\section{LED fabrication}

\begin{figure}[t]
\centering
\includegraphics{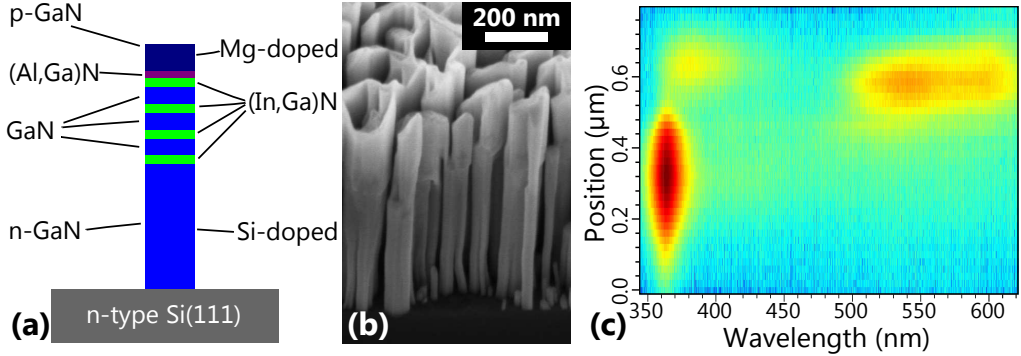}
\caption{(a) Schemativ view of the NW LED structure and (b) SEM bird's-eye view image of the NW ensemble after MBE growth and prior to processing. (c) CL spectral linescan acquired at room temperature along the center of a single dispersed NW. The position indicates the distance to the NW base end. The CL intensity is color-coded on a logarithmic scale and increases from blue via green and yellow to red.}\label{fig1}
\end{figure}

NW LEDs were fabricated by embedding heterostructures that consist of (In,Ga)N multiple QWs (MQWs) in the intrinsic region of GaN NWs with an n-i-p diode doping profile. The NWs were grown on Si(111) substrates by plasma-assisted MBE in a self-induced way, i.\,e. without any external catalysts or pre-patterning \cite{Sanchez-Garcia1998, Debnath2007, Geelhaar2011, Consonni2011b}.
In figure~\ref{fig1}(a), a schematic view  of the NW heterostructure is presented along with a scanning electron microscope (SEM) image of the as-grown NW ensemble in figure~\ref{fig1}(b). Special care was taken so that the NW morphology was not affected by the supply of Si during the growth of the Si-doped GaN base. The doping of NWs is a general challenge, as the complex growth mechanisms can be significantly altered by the addition of dopants \cite{Furtmayr2008, Furtmayr2008a, Richter2008, Jeganathan2009, Limbach2010, Limbach2012, Stoica2011}. In separate growth experiments, we determined that the maximum Si flux for which the nanowire morphology is not significantly changed corresponds to a nominal Si concentration of $3 \times 10^{19}$\,cm$^{-3}$ \cite{SIMS}. This Si flux was used for the growth of the NW base. After the growth of the GaN:Si base at 780\,$^\circ$C, the substrate temperature was lowered to 604\,$^\circ$C, in order to enable the incorporation of In for the formation of QWs. The MQW structure consists of four wells with an In content of ($20\pm10$)\% and a width of ($3\pm1$)\,nm, determined by x-ray diffraction on reference NW samples \cite{Wolz2011,Kaganer2011}. The barriers were designed to have a thickness of 8\,nm. The last QW was immediately followed by a Mg-doped Al$_{0.15}$Ga$_{0.85}$N electron blocking layer with a nominal thickness of 13\,nm. Finally, a Mg-doped GaN cap layer was grown. Again, precautions were taken to prevent the doping to destroy the NW morphology. A very low growth rate of only 0.8\,$\textrm{nm}/\textrm{min}$ was employed to limit coalescence of the NWs as far as possible. For the same reason, the growth temperature was gently raised to 744\,$^\circ$C at 2\,$^\circ\textrm{C}/\textrm{min}$ during the initial phase of the cap growth. Nevertheless, the SEM image in figure~\ref{fig1}(b) indicates that some coalescence took place, and it seems rather challenging to incorporate Mg without any effect on the NW morphology \cite{Limbach2010, Furtmayr2008, Furtmayr2008a, Stoica2011}.

In order to verify the growth of the designed LED structure, spectrally and spatially resolved cathodoluminescence (CL) measurements were performed.
Figure~\ref{fig1}(c) shows a CL spectral linescan acquired at room temperature (RT) along a single dispersed NW with a resolution  of the spectrometer of $\approx3$\,nm. All three parts of the LED structure can be identified. The bottom part with a hight of 500\,nm are dominated by the near band edge (NBE) luminescence of GaN at 364\,nm as expected for n-type GaN \cite{Reshchikov2005}. This segment is followed by the MQW structure and its broad luminescence band around 560\,nm \cite{Limbach2011,Lahnemann2011, Wolz2012}. Finally, the Mg-doped GaN cap with a length of about 150\,nm can be identified by a redshift of the luminescence compared to the n-type region \cite{Limbach2010, Furtmayr2008a, Reshchikov2005}. This redshift indicates donor-acceptor pair transitions caused by incorporated Mg. Overall, the CL measurements confirm that the NW LED structure was grown as designed.

\begin{figure}[t]
\centering
\includegraphics{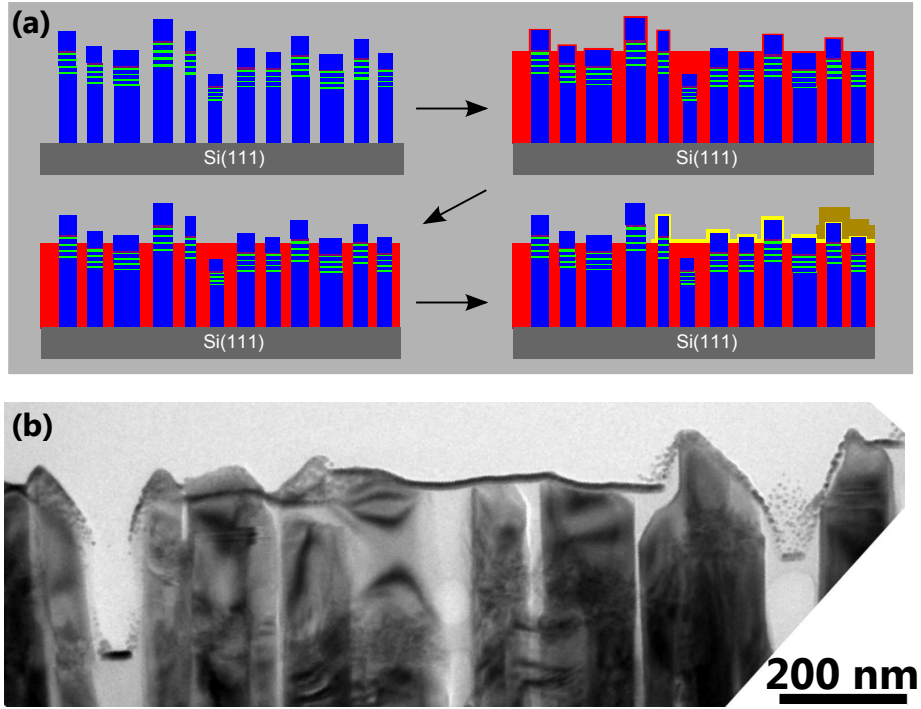}
\caption{(a) Sketches of the sample geometry during processing, i.e.\ as grown, planarized, after etching, and with semi-transparent front contact and contact pads. (b) Cross-sectional TEM image of the processed NW LED.}\label{fig2}
\end{figure}
The different stages of the LED processing are sketched in figure~\ref{fig2}(a). The NW ensemble was planarized by spin coating a solution of hydrogen silsesquioxane (HSQ), which was subsequently transformed into amorphous SiO$_{\textrm{x}}$. The SiO$_{\textrm{x}}$ acts as an insulator between the NWs. In order to uncover the Mg-doped GaN NW tips, a back-etching step is necessary. Next, the Mg-doped NW tips were contacted by i) a semi-transparent front metallization (Ni/Au, 5\,nm\,/\,5\,nm)  and ii) contact pads for bonding and current spreading (Ti/Au, 10\,nm\,/\,90\,nm). Finally, the backside of the Si substrate was metallized with Al/Au (50\,nm\,/\,50\,nm) to form the n-type contact.

A particular challenge for the processing of such an LED based on a NW ensemble consists of the fluctuation of the NW height. Thus, the metal layer on top may not be continuous, and short NWs may actually not be contacted. In order to investigate the NW top contact, the LEDs were analyzed by cross-sectional transmission electron microscopy (TEM). In the image presented in figure~\ref{fig2}(b) the semi-transparent contact is visible as the dark edge. This image demonstrates, in particular, that the two contact layers of Ni and Au are well connected from NW to NW. Also, Ni is in direct contact with the NW tips, as desired. Both goals were achieved for the majority of the NWs (the image shown is representative for several areas investigated by TEM). 

\section{Basic LED characterization}

\subsection{Current-voltage characteristics}

\begin{figure}[t]
\centering
\includegraphics{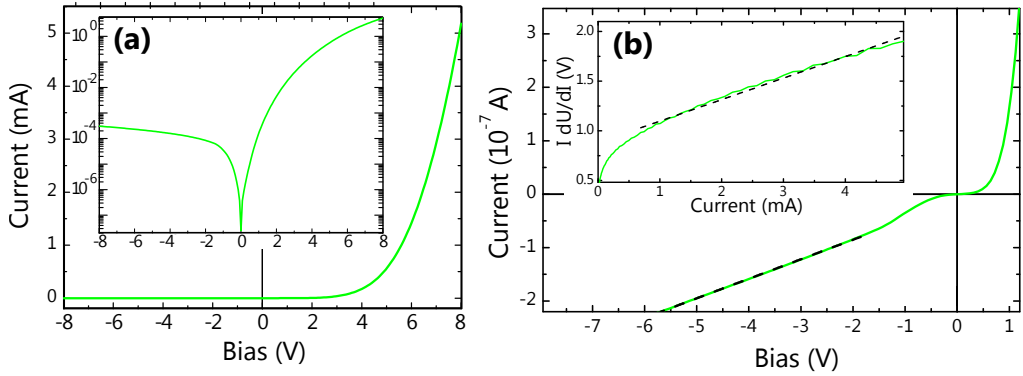}
\caption{(a) Current-voltage characteristics of a processed LED with a contact area of 0.2\,mm$^2$ on a linear and (inset) logarithmic scale. (b) Current-voltage characteristics for small currents, the dashed line is the linear fit for the determination of $R_p$. In the inset, the product of current $I$ and the derivative of the voltage $U$ with respect to $I$ is plotted against $I$. The dashed line in the inset is the linear fit from which $R_S$ is determined using only data points for which $U \gg k_B T / e$.}\label{fig4}
\end{figure}

Current-voltage measurements for a contact pad of 0.2\,mm$^2$ are presented in figure~\ref{fig4}(a) on a linear and logarithmic (inset) scale. The turn-on voltage of the device is approximately 5.2\,V. The exact determination of this value is not possible as each individual NW in the contacted ensemble constitutes an LED with its own turn-on voltage (see also section 4.2). Therefore, the value of 5.2\,V is only an average of turn-on voltages for the NWs active at 8\,V. The high value of the turn-on voltage is in line with earlier reports on similar devices \cite{Guo2010, Armitage2010, Bavencove2011} and can in part be explained by the amorphous Si$_{\textrm{x}}$N$_{\textrm{y}}$ layer between the NWs and the Si substrate \cite{Geelhaar2011, Calleja2007, Stoica2008}. In addition, since the processing is not as mature as for conventional LEDs, also the contacting might contribute significantly to the high turn-on voltage. In contrast, the leakage current of 1.5\,$\times$\,10$^{-6}\,\textrm{A}/\textrm{mm}^{2}$ at $-$8\,V indicates a very good insulation of the individual NWs by the SiO$_{\textrm{x}}$ and is two orders of magnitude lower than what has been published previously \cite{Kikuchi2004, Hersee2009, Guo2010, Kunert2011, Bavencove2011}.

Plotting the small-current range of the characteristics in an appropriate scale as shown in figure \ref{fig4}(b) reveals a non-vanishing slope in the reverse bias regime and therefore indicates the presence of a parallel current path with resistance $R_P$. Its value can be determined by a linear fit to the reverse bias regime \cite{Schubert2006}. Such a fit yields a parallel resistance of $R_P= (27.3 \pm 0.1)\,$M$\Omega$ which is an excellent value for this device compared to the data published so far \cite{Kikuchi2004, Hersee2009, Guo2010, Kunert2011, Bavencove2011}.

In the forward bias regime, the current-voltage characteristics deviate from the ideal behaviour described by the Shockley equation as well. This deviation is in part caused by a series resistance $R_S$ in the current path \cite{Schubert2006} due to both the Si$_{\textrm{x}}$N$_{\textrm{y}}$ interlayer present at the bottom of the GaN NWs \cite{Calleja2007, Stoica2008} and the contacts at the top. The inset of figure~\ref{fig4}(b) depicts a plot of $I (dU/dI)$ versus $I$, and the value of $R_S$ can be determined from its slope \cite{Schubert2006}. However, the slope is not perfectly linear due to the tunneling resistance of the Si$_{\textrm{x}}$N$_{\textrm{y}}$ interlayer which decreases at high bias voltages. Therefore, a large error for $R_S$ is associated with this approach. The linear fit to the slope results in a series resistance of $R_S  = (220 \pm 40)\,\Omega$. This rather large value contributes to the high turn-on voltage that presently seems to be typical for this kind of LED \cite{Guo2010, Armitage2010, Bavencove2011} and indicates room for improvement in the device processing.

\subsection{Electroluminescence}

\begin{figure}[t]
\centering
\includegraphics{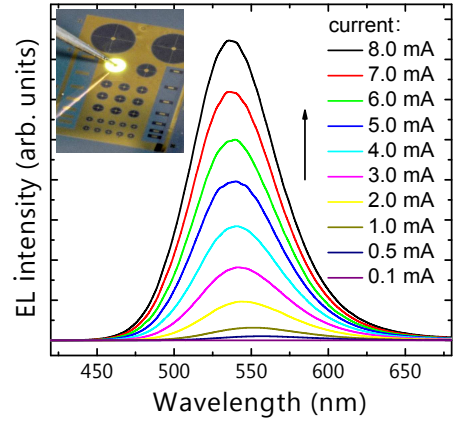}
\caption{EL spectra of the NW LED at room temperature for different forward currents. The inset shows a photograph of such a device with a diameter of 1\,mm under 8\,V forward bias.}\label{fig5}
\end{figure}

For preliminary EL measurements, the devices were contacted on a probe station. The inset of figure~\ref{fig5} shows a photograph of a NW LED with a diameter of 1\,mm in operation. This picture demonstrates the efficient current spreading through the semi-transparent contacts for this device. The EL can be seen with the naked eye at around 4\,V. The main part of figure~\ref{fig5} shows the current-dependent EL spectra of the ensemble NW LED at room temperature. No emission from GaN is visible (wavelength region not shown) which indicates a very good overlap of the p-n-junction with the active zone. The EL emission is centered around 540\,nm in agreement with the CL emission from the (In,Ga)N QW region shown in figure \ref{fig1}. The full width at half maximum of the EL is 68\,nm at 8\,mA which is typical for this kind of LED \cite{Kikuchi2004, Guo2010, Bavencove2011, Lin2010}. The rather large value suggests inhomogeneity in the emission wavelength of the individual NW LEDs, since the presented ensemble measurement integrates all spectra of the individual NWs \cite{Lahnemann2011}. The slight shift in the peak position with increasing current is consistent with the occurrence of the QCSE that will be addressed in section 5.

\begin{figure}[t]
\centering
\includegraphics{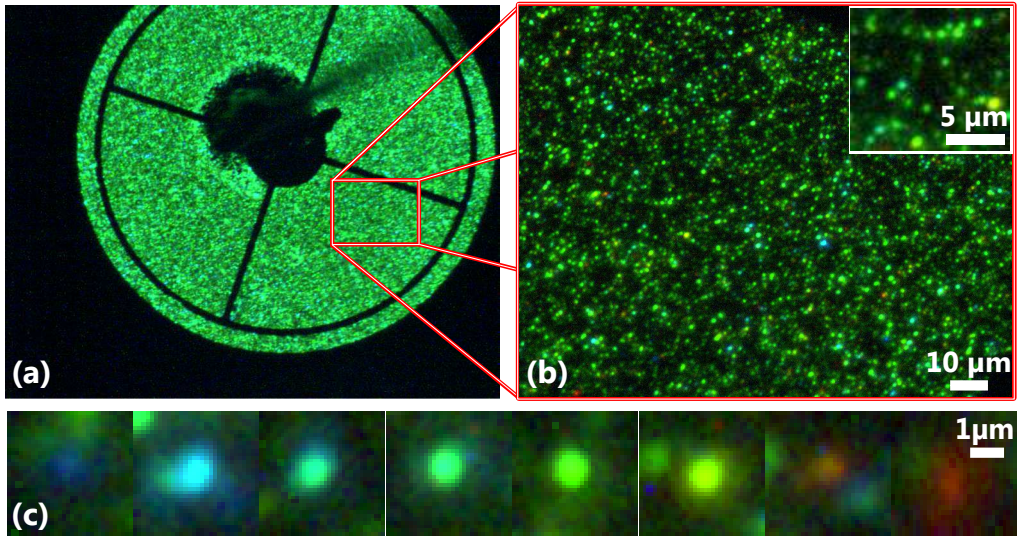}
\caption{EL of a NW-LED with an area 0.2\,mm$^2$ under 10\,V forward bias imaged through a microscope at (a) 200$\times$ and (b) 500$\times$ magnification. The inset in (b) shows an enlarged part of the image. The lower part (c) shows  close-ups of individual luminescence spots taken from the image with 500$\times$ magnification.}\label{fig6}
\end{figure}

In order to clarify the inhomogeneity, EL maps were acquired at 200 and 500$\times$ magnification as presented in figures~\ref{fig6}(a) and (b). In agreement with previous findings \cite{Bavencove2011}, the NW LED exhibits a very spotty emission pattern. One can find emission spots with different colors ranging from blue all the way to red as demonstrated in the close-up images in figure~\ref{fig6}(c), while the vast majority of spots exhibits green emission. The diameter of the emission spots is diffraction limited, so that it is plausible that each spot corresponds to a single NW emitting light.

The total number density of these electroluminescent spots is around $1\times 10^{7}\,\textrm{cm}^{-2}$ at 10\,V, i.\,e. roughly two orders of magnitude lower than the NW density on the sample after growth \cite{Limbach2012, Calarco2007, Consonni2011a}. Similar observations of such a low density of electroluminescent spots have been reported recently, suggesting that this phenomenon is typical for LEDs based on NW ensembles \cite{Kishino2007, Lin2010,Bavencove2011}. Moreover, isolated electroluminescent spots as seen on the inset in figure~\ref{fig6}(b) would not be observed if a significant fraction of NWs would emit light. Obviously, the fact that only a small fraction of NWs emits EL is unacceptable for applications. The origin of this phenomenon is elucidated in the next section.

\section{Current path and quantum efficiency}

The large number of dark NWs may be caused by a negligible current injected into the majority of NWs or by a very low internal quantum efficiency (IQE). A low IQE could be explained by growth-induced defects whose concentration fluctuates between individual NWs. An inefficient current injection could, in principle, result from several factors: First, the height distribution of the NWs and the planarization of the ensemble may lead to the burial of some NWs that are significantly shorter than the mean. Second, for those NWs which are significantly longer than the mean the metallization of the MQW structure or even the n-type base is possible. Third, a locally higher thickness of the Si$_{\textrm{x}}$N$_{\textrm{y}}$ layer may significantly increase the series resistance for some NWs. Fourth, variations in the metallization may also result in locally higher contact resistance. 

The latter two of these factors relate to fluctuations in the series resistances, and such fluctuations were indeed already implied by the observation that the ensemble LED does not exhibit a well-defined turn-on voltage (cf. section 3.1). We investigated this phenomenon in more detail by recording EL maps for various forward biases as discussed in section 4.1. The first two of the above factors seem unlikely to be important in the present case since TEM indicates a rather homogeneous height distribution and continuous metallization [figure \ref{fig2}(b)]. Nevertheless, a merely structural characterization does not prove the realization of an electrical contact. Thus, in section 4.2 we describe experiments that assess directly the contacting of the NWs. In order to determine whether there is indeed a significant variation in IQE between the NWs, we measured their luminescence independently of the current path, and these results are presented in section 4.3. 

\subsection{Bias-dependence of spot density}

\begin{figure}[t]
\centering
\includegraphics{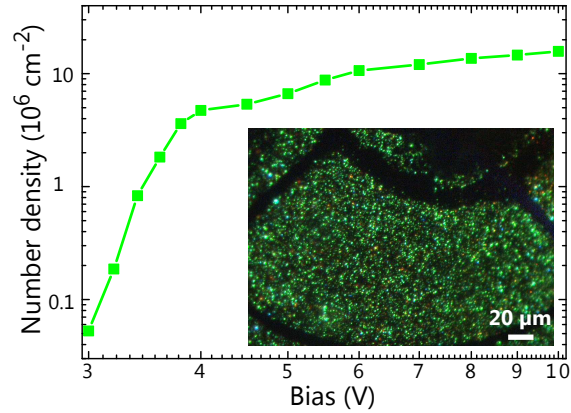}
\caption{Number density of electroluminescence spots as a function of the applied forward bias. The inset is a screen shot of a movie showing the EL during a ramp from 3\,V to 10\,V forward bias. The data were extracted from the central region of the individual frames. The movie is provided as supplementary data.}\label{fig7}
\end{figure}

In order to analyze variations in the series resistances of the NWs in more detail, we acquired EL maps for various forward biases. The extracted number densities of electroluminescent spots are plotted in figure~\ref{fig7}, and the corresponding images have been converted into a movie (supplementary). Already at 3.4\,V a significant number of NWs emit light, i.\,e. well below the turn-on voltage found for the device as a whole, and the number density of NWs showing electroluminescence increases monotonically up to 10\,V. 
Both observations indicate that there is indeed a significant inhomogeneity of the series resistances: The LED has to be considered as an ensemble of many individual NW-LEDs contacted in parallel rather than a uniform device. Each of the NWs exhibits an individual $R_S$ and turn-on voltage, as also discussed by Bavencove \textit{et al.} \cite{Bavencove2011}. In addition, the series of EL maps reveals that the increase in ensemble luminescence intensity with increasing current seen in figure~\ref{fig5} is caused by an increase both in the number density of actually luminescent NWs and in emission intensity from individual NWs. 

For the interpretation of these measurements, one has to take into consideration that they reveal information only about those NWs that eventually do emit light. However, these data do not account for the large number of NWs that remain dark, so it is still mandatory to assess contacting and IQE.

\subsection{Electrical contacts}

\begin{figure}[t]
\centering
\includegraphics[width=13.5cm]{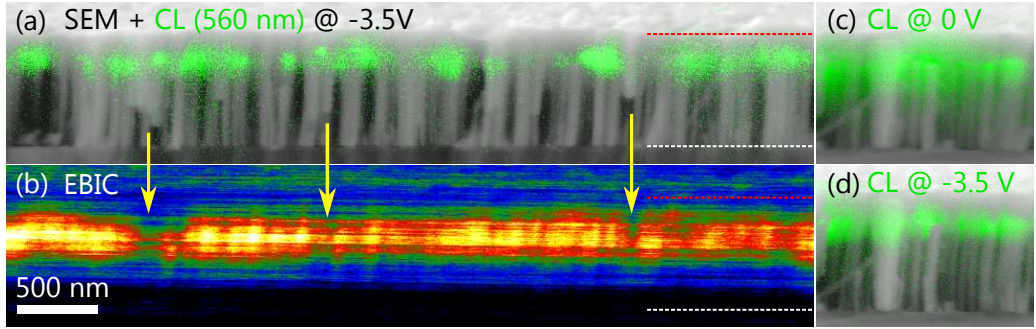}
\caption{(a) Monochromatic room temperature CL image recorded around 560\,nm with a bandpass of approximately 30 nm superimposed on an SEM image and (b) corresponding EBIC map with color-coded intensity (increasing from black via blue, green, red, and yellow to white). These measurements on the sample cross section were performed under a reverse bias of $-$3.5~V. The dashed horizontal lines indicate the substrate-NW (white) and NW-top-contact (red) interface, respectively. The yellow arrows highlight broken NWs that do not contribute to the EBIC signal. (c) and (d) CL images of the same region acquired unbiased and under reverse bias, respectively.}
\label{fig3}
\end{figure}

Conclusive evidence that most NWs were successfully contacted can be derived from electron-beam-induced current (EBIC) measurements. To this end, a processed LED structure was cleaved and contacted in order to perform measurements on a cross-section of the device. Our set-up allows the simultaneous acquisition of secondary-electron (SE), CL and EBIC signals. Such a set of measurements is depicted in figure~\ref{fig3}. 
The monochromatic CL of the (In,Ga)N band around 560~nm is superimposed on the corresponding SE image in figure~\ref{fig3}(a), which visualizes the position of the QWs along the NW cross-section and shows that the top contact is well above the QW region, despite the variation in NW length. Note, that most NWs exhibit CL although the intensity varies significantly. The latter two aspects will be analyzed in more detail in subsection 4.3 with the help of top-view CL images.

The EBIC map associated with the same SE image is depicted in figure~\ref{fig3}(b). In EBIC, electron-hole pairs created in or diffusing to the depletion region of the p-n-junction are separated by the electric field. The resulting short-circuit current can be detected through an external current amplifier while the electron beam is scanned across the sample. In figure~\ref{fig3}(b), the bright stripe in the middle of the NWs indicates the position of the p-n junction. An additional, but weaker, EBIC signal can be detected from the region of the top contact implying the presence of a slight band bending at the semiconductor-metal interface. The strong EBIC signal related to the p-n junction directly shows which NWs are contacted. In fact, those NWs not contributing to the EBIC signal in figure~\ref{fig3}(b) are broken and thus not connected to the back contact (yellow arrows). A reverse bias of $-3.5$\,V was applied since the EBIC signal is too weak at 0\,V. The reverse bias increases the width of the depletion region of the diode and thus reduces the amount of carriers recombining in the QWs, while increasing the one contributing to the EBIC signal. At an acceleration voltage of 8~kV for the electron beam, most of the signal originates from the first row of NWs. Therefore, the cross-sectional EBIC map directly visualizes that the majority of these NWs is contacted.

Carrier trapping by the QW and carrier separation by the p-n-junction are competing processes after excitation by the electron beam. Under reverse bias, the drift of carriers to the contacts induced by the increased electric field dominates over the diffusion to the (In,Ga)N QWs. Therefore, only carriers excited directly at the QWs contribute to the CL signal and the spatial resolution of the CL is thus improved as can be seen in the comparison of the CL images in figure~\ref{fig3}(c) and (d) that were acquired at a bias of 0\,V and $-3.5$\,V, respectively. In the first case, electrons and holes can diffuse to the MQW and recombine radiatively even if they are excited outside the active region. In contrast, under reverse bias the CL signal is only recorded when the electron beam directly excites the QWs. Therefore, figure~\ref{fig3}(a) precisely reflects the position of the QWs in the NWs. This position coincides with the upper part of the depletion region visualized in the EBIC map.

\subsection{Internal and external quantum efficiencies}

\begin{figure}[t]
\centering
\includegraphics{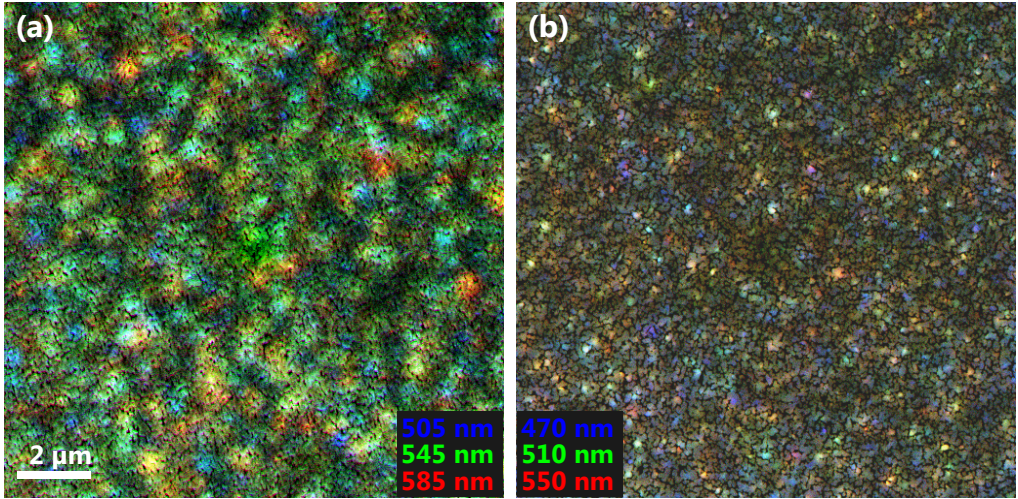}
\caption{Top-view false-color CL images taken at room temperature of (a) the NW-LED sample (unprocessed) and (b) an undoped reference sample. To cover the broad (In,Ga)N contribution, a spectral bandpass of approximately $50$\,nm was chosen and three monochromatic images were superimposed which were recorded at the peak wavelength (green), and on the short (blue) and long (red) wavelength flanks. Additional colors result from an overlap of the monochromatic images. The bandpass regions were slightly adjusted for the two samples to accommodate minor differences in the MQWs. No SE image is superimposed.}
\label{fig8}
\end{figure}

To assess the homogeneity of the luminescence independent of the current path, a top-view CL image of the as-grown NW ensemble has been recorded with a wide spectral bandpass of about $50$\,nm as shown in figure~\ref{fig8}(a). This false-color image is dominated by luminescent spots with a diameter of 200--500\,nm. The spots indicate luminescence centers which collect carriers excited by the electron beam. The number density of these spots of about $1\times 10^{8}\,\textrm{cm}^{-2}$ is one order of magnitude higher than observed in EL, but still one order of magnitude lower than the NW density. The size of the spots is partly a result of the electron beam interaction volume at an acceleration voltage of 8\,kV, chosen to penetrate the cap layer and excite the QWs. However, the major effect for this spatial broadening is carrier diffusion within the partially coalesced p-type cap segment [cf.\ figure~\ref{fig1}(b)], i.\,e. between neighboring NWs. Carriers excited in the cap may diffuse along local minima in the potential landscape to QWs even in neighboring NWs and recombine radiatively there rather than at the position of the electron beam during their excitation. Recombination centers emitting at lower energies (green and red) typically collect carriers from a larger area and thus the spots have a larger diameter than for higher energies (blue). Hence, the CL image presented in figure~\ref{fig8}(a) is largely affected by the relaxation of carriers into potential minima and does not reflect the capability of the individual NWs to emit light.

The coalescence leading to the carrier diffusion between neighboring NWs is a result of the Mg doping. Therefore, an undoped reference sample with similar QWs emitting at a slightly shorter wavelength, but with a similar PL intensity under resonant excitation has also been investigated. It is shown in figure~\ref{fig8}(b), the spot diameter decreases with the reduced degree of coalescence and the total number of luminescence spots agrees fairly well with the NW number density. This result shows conclusively that the low percentage of emitting NWs in the EL image of figure~\ref{fig6} is not caused by a significantly reduced IQE for the majority of the NWs.

At the same time, the emission intensity of a few NWs in the reference sample is significantly higher than the mean, and their number density is similar to the one observed in figure~\ref{fig8}(a). Which may be atributed to differences in the IQE of individual NWs. Of course, such differences in the IQE of individual NWs will occur in a self-induced NW ensemble, in which fluctuations of the QW thickness from NW to NW are essentially inevitable. However, an effect just as inevitable for a random array of NWs is the fluctuation of the extraction efficiency. Whether we view the NW ensemble as a disordered photonic crystal in which multiple light scattering contributes to light extraction \cite{Long2008,Yang2008} or as an inhomogeneous effective medium in the limit of very small NW dimensions and distances \cite{Asatryan1999} does not change the result: the spatially random arrangement of dielectric cylinders results in areas of incidentally enhanced extraction efficiency. A closely related subject is the random lasing observed upon optical pumping for GaN NW ensembles in which spatial light localization occurs by chance \cite{Sakai2010}.

\subsection{Discussion}

Taking all the results of this section into account, we can now analyze the current path in LEDs based on NW ensembles. The TEM characterization [figure~\ref{fig2}(b)] and the EBIC investigation [figure~\ref{fig3}(b)] demonstrate that the vast majority of NWs is electrically contacted. However, the electrical characteristics (figure~\ref{fig4}) and the voltage-dependent EL maps (figure~\ref{fig7}) show that the turn-on behaviour of the individual NWs varies significantly. Moreover, the low number density of isolated, well-defined emission spots on EL images [figure~\ref{fig6}(b)] implies that many NWs do not emit EL at all. At the same time, our CL investigation [figure~\ref{fig3}(a) and figure~\ref{fig8}(b)] indicates that independent of the current path essentially all the NWs emit light, yet they exhibit fluctuations in intensity. For EL emission, an individual NW has to have a sufficiently high IQE for light emission and has to carry current, so both types of inhomogeneities have to be considered. The IQE of a given NW is not affected by the IQEs of the neighboring NWs. In contrast, NWs carrying a high current density have to be surrounded by NWs with low current density because the two values are not independent. Hence, the consequences of fluctuations in individual series resistance and thus local current density are much more severe than those of variations in IQE, in agreement with our experimental results. Therefore, we conclude that the widely differing current injection into the individual NW LEDs determines the EL pattern of an LED based on a NW ensemble.

\section{NW polarity and QCSE}

\begin{figure}[b]
\centering
\includegraphics{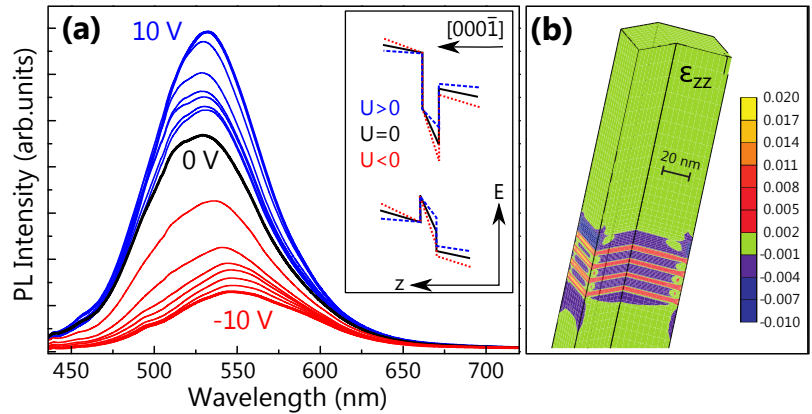}
\caption{(a) $\mu$-PL spectra of the processed NW LED recorded at room temperature for an applied voltage U in forward (blue) and reverse (red) direction. For forward voltages, the EL spectra of the LED were measured separately and subtracted from the PL spectra. The inset shows a schematic representation of the band profile of a N-polar (In,Ga)N/GaN QW under the influence of an external voltage U. (b) The out-of-plane strain within the NW as obtained by finite element simulations is depicted.}\label{fig9}
\end{figure}

Processed NW LEDs offer the possibility to study the PL of the NWs while applying a bias. Such an investigation enables us to obtain additional important insight about the (In,Ga)N/GaN NWs, as we will describe in this section. Figure~\ref{fig9}(a) shows the $\mu$-PL spectra recorded at room temperature under resonant excitation with the 413.1\,nm line of a Kr$^{+}$ laser, a spot size of about $1$\,$\mu$m and a total beam power of 8\,mW. The NWs show a broad (In,Ga)N-QW emission centered around 530\,nm under biased as well as unbiased conditions as observed also in EL (see figure~\ref{fig5}). For positive voltages the EL, which contributes only 8\,\% to the total signal at 10\,V and is slightly redshifted compared to the bare PL signal, was measured independently and subtracted from all $\mu$-PL spectra. Applying a forward bias to the LED leads to a moderate but steady increase of the PL intensity and no significant shift in transition energy up to $+10$\,V. In reverse direction, the PL intensity is quenched significantly while the spectrum shifts abruptly to lower transition energies. The fact that the PL intensity is significantly influenced by the applied bias confirms that most of the NWs are contacted electrically, since all (In,Ga)N-QWs are excited simultaneously in this experiment.

To investigate the possibility of carrier escape from the QWs under reverse bias, we also recorded the photocurrent during the biased PL experiment. At a reverse bias of -2\,V, where the PL signal is already significantly quenched, a photocurrent of only -22\,$\mu$A is measured. At -10\,V the photocurrent reaches $-100$\,$\mu$A. Comparing those values to the current under forward bias in figure~\ref{fig4}(a), it becomes immediately clear that carrier escape from the QWs cannot be responsible for the quenching of the PL under reverse bias.

The observed behavior of the PL intensity reflects that the external bias changes the strength of the electric field within the QWs. An increase of this field, for example, reduces the electron-hole overlap and thus the internal quantum efficiency, which in turn leads to a lower PL intensity. The electric field results from the superposition of the built-in depletion field of the p-i-n-junction, the internal electrostatic field primarily induced by the piezoelectric polarization of the (In,Ga)N QWs, and the applied voltage. For a comparison of these contributions to the total electric field, we utilize finite element simulations of the strain in the (In,Ga)N QWs embedded in the GaN NW. Figure~\ref{fig9}(b) shows the component $\varepsilon_{zz}$ of the elastic strain tensor. The out-of-plane strain reaches a value of 1\% in the center part of the QW, which is close to the value expected for a laterally infinite planar (In,Ga)N QW (1.2\%). This strain within the QWs results in a piezoelectric polarization $P_z = 0.02$~ C/m$^2$. Self-consistent Schrödinger-Poisson calculations of the band profile of the structure along the NW axis show that this polarization induces an internal electrostatic field of 2~MV/cm, which is significantly stronger than the built-in depletion field of the p-i-n-junction. Assuming electron and hole concentrations in the n- and p-type segments of $10^{18}$ and $10^{17}$~cm$^{-3}$, respectively, this built-in field within the intrinsic segment of the NW amounts to 300~kV/cm.        

This estimate of the internal fields allows us to draw an important conclusion about the polarity of the NWs. The dependence of the PL intensity upon the applied bias implies that the total field in the QWs is reduced under forward but increased under reverse bias. Together with the fact that the field due to the internal piezoelectric polarization significantly exceeds the built-in field of the p-i-n-junction, this specific dependence of the total field on bias is only possible when the NWs exhibit N polarity, i.~e., are grown along the $[000\overline{1}]$ direction. The resulting band profile is schematically shown in the inset of figure~\ref{fig9}(a). Note that the field within the QW would change its sign for Ga polar NWs, and the dependence of the electron-hole overlap and thus of the PL intensity on the external field would be reversed. It is well known that self-induced GaN NWs grow in MBE along the c-axis \cite{Geelhaar2011}, but conflicting results have been published with respect to the NW polarity \cite{Hestroffer2011, Brubaker2011, Cherns2008, Geelhaar2011, Kong2011, Mata2012}. Our experiments support the recent and comprehensive report which indicates that these NWs are N-polar by Hestroffer et al. \cite{Hestroffer2011}. 

Finally, we address the QCSE. It has been argued that (In,Ga)N QWs in such NWs are completely strain-free and thus not subject to the QCSE \cite{ Nguyen2011, Armitage2008, Guo2011}, but there are also contradicting publications \cite{Lahnemann2011, Bavencove2011}. Here, we find in agreement with our previous study \cite{Lahnemann2011} that the (In,Ga)N QWs still experience the QCSE.

\section{Summary and conclusions}

The EL of LEDs based on NW ensembles is characterized by a spotty emission pattern because such devices have to be considered as arrays of many tiny LEDs operated in parallel. In such LEDs frequently only a small fraction of the NWs emits light under typical biases. We identified that this drawback is caused mainly by spatial inhomogeneities in the current path. While there are also variations in IQE of the individual NWs, these variations are by far not as pronounced. The NW-to-NW fluctuations in series resistances must be significant and lead to fluctuations in current density between the single-NW LEDs, and only those NWs with high current density emit EL.

In addition, our PL measurements under bias support the recent findings that self-induced GaN NWs on Si grow in the N-polar direction\cite{Hestroffer2011} and that the (In,Ga)N QWs in such NWs still experience the QCSE \cite{Lahnemann2011}.

For the future, a more homogeneous NW ensemble is desirable in order to reduce the inter-NW differences in series resistance. This may be achieved by the usage of selective-area grown NWs for LED applications since it is expected that the NW-to-NW fluctuations can be reduced significantly by this approach \cite{Sekiguchi2010,Bertness2010,Gotschke2011,Schumann2011a, Li2012}. While we analyzed the specific case of an LED based on an (In,Ga)N/GaN NW ensemble grown by MBE, our conclusion about the need for homogeneity in current path holds true for all devices based on NW ensembles and can therefore be transferred to other growth techniques, other material systems, and even other types of devices.

\newpage

\bibliographystyle{iopart-num}
\bibliography{lit}

\end{document}